\documentclass[letterpaper,english,reprint, aps]{revtex4-1}
\usepackage[T1]{fontenc}
\usepackage[latin9]{inputenc}
\usepackage{booktabs}
\usepackage{units}
\usepackage{textcomp}
\usepackage{amsmath}
\usepackage{amssymb}
\usepackage{graphicx}

\makeatletter

\providecommand{\tabularnewline}{\\}

\makeatother

\usepackage{babel}
\begin{document}
\title{Conductivity exponents at the percolation threshold}
\author{Clinton DeW. Van Siclen}
\email{cvansiclen@gmail.com}

\address{1435 W 8750 N, Tetonia, Idaho 83452, USA}
\date{4 January 2021}
\begin{abstract}
Connections are found between the two-component percolation problem
and the conductor/insulator percolation problem. These produce relations
between critical exponents, and suggest formulae connecting the conductivity
exponents in different dimensions. Values for the critical exponents
are obtained from calculations on the incipient infinite cluster in
two and three dimensions.
\end{abstract}
\maketitle

\section{Introduction}

Percolation is a prototypical example of a critical phenomenon \citep{SA}.
In particular, a percolating system is characterized by a correlation
length $\xi$ that diverges as the percolation threshold $p_{c}$
is approached. At the critical point $p_{c}$, the geometric and dynamic
attributes of the infinite, percolating cluster (termed the \textquotedblleft incipient
infinite cluster\textquotedblright ) are identified with a set of
critical exponents whose values collectively constitute a universality
class; that is, the set of exponent values is particular to the dimension
of the Euclidean space rather than the underlying (regular) lattice
structure. Because microscopic details of the system near $p_{c}$
are not important, percolation serves as a useful model for a variety
of natural phenomena \citep{Sahimi} where a dynamical process is
affected by the connectivity of the system. In this paper, however,
uncorrelated, isotropic systems are considered, where the interest
is in the values of the critical exponents and the relations between
them.

Two different approaches to the critical point (at which $\xi$ becomes
infinite) are taken by the two-component percolation problem and the
more-familiar conductor/insulator percolation problem. These two systems
have no geometric attributes in common, but are related by their dynamic
exponents.

The two-component percolation problem \citep{Straley} involves a
two-component material system where the higher conductivity phase,
having conductivity $\sigma_{1}$, is randomly mixed with the lower
conductivity phase ($\sigma_{2}$); further, the volume fraction $p$
of the higher conductivity phase is precisely at the percolation threshold
$p_{c}$. It is reasonable to expect the effective conductivity $\sigma$
of the system to exhibit critical behavior as the conductivity value
$\sigma_{2}$ approaches zero. Indeed, the power-law relation
\begin{equation}
\sigma=\sigma_{1}\,r^{u}\label{eq1}
\end{equation}
where ratio $r\equiv\sigma_{2}/\sigma_{1}<1$, is found to hold for
2D square bond \citep{Straley} and square site \citep{VS99} lattice
systems of size greater than $\xi(r)$. The correlation length $\xi$
is the length scale above which the property values (in this case
the effective conductivity) of the \textit{infinite} system are obtained.

The conductor/insulator percolation problem involves an insulator
phase randomly mixed with a conducting phase of volume fraction $p>p_{c}$.
The effective conductivity $\sigma$ exhibits the \textit{asymptotic}
behavior
\begin{equation}
\sigma\sim(p-p_{c})^{t}\label{eq2}
\end{equation}
as $p$ approaches $p_{c}$ from above. In this case the correlation
length $\xi(p)\rightarrow\infty$ as $p\rightarrow p_{c}$, so the
value of the critical exponent $t$ is obtained from the infinite
system at $p=p_{c}$.

{[}Some comments on notation: The tilde symbol $\sim$ indicates that
the quantities are related by similar asymptotic behavior (as in the
equation above). The letter $t$ is used both for the conductivity
exponent (as in the equation above) and for the variable \textquotedblleft time\textquotedblright ;
it should be clear from the context, and placement, what meaning should
be assumed for $t$. In parts of this paper it is convenient to denote
an effective conductivity in a more particular way than is done above.
For example, $\sigma(p,\sigma_{1};(1-p),\sigma_{2})$ is the effective
conductivity of an uncorrelated system comprised of volume fraction
$p$ of sites having conductivity $\sigma_{1}$, and volume fraction
$(1-p)$ of sites having conductivity $\sigma_{2}$.{]}

The following section presents the Walker Diffusion Method by which
the analytical and numerical results in this paper are obtained. Subsequent
sections are devoted to the two-component percolation problem, the
conductor/insulator percolation problem, and numerical methods and
results. An appendix extends the WDM to bond-based percolating systems.

\section{Walker Diffusion Method}

The WDM was developed to calculate effective transport coefficients
(e.g., conductivity) of composite materials and systems \citep{VS99}.
This method exploits the isomorphism between the transport equations
and the diffusion equation for a collection of non-interacting walkers
(hence the name). Accordingly, the phase domains in a composite correspond
to distinct populations of walkers, where the walker density of a
population is given by the value of the transport coefficient of the
corresponding phase domain. The principle of detailed balance ensures
that the population densities are maintained, and provides the following
rule for walker diffusion over a digitized (pixelated) composite:
a walker at site (or pixel) $i$ attempts a move to a randomly chosen
adjacent site $j$ during the time interval $\tau=(4d)^{-1}$, where
$d$ is the Euclidean dimension of the space; this move is successful
with probability $p_{ij}=\sigma_{j}/(\sigma_{i}+\sigma_{j})$, where
$\sigma_{i}$ and $\sigma_{j}$ are the transport coefficients for
the phases comprising sites $i$ and $j$, respectively. (In practice,
the unsuccessful moves inherent in this rule are eliminated by use
of the variable residence time algorithm \citep{VS99}.) The path
of a walker thus reflects the composition and morphology of the domains
that are encountered. Over distances greater than the correlation
length $\xi$, the walker diffusion is characterized by the diffusion
coefficient $D_{w}$, which is related to the effective transport
coefficient $\sigma$ by
\begin{equation}
\sigma=\left\langle \sigma(\mathbf{r})\right\rangle D_{w}\label{eq3}
\end{equation}
where $\left\langle \sigma(\mathbf{r})\right\rangle $ is the volume
average of the constituent transport coefficients. The diffusion coefficient
$D_{w}$ is calculated from the equation
\begin{equation}
D_{w}=\frac{\left\langle R(t)^{2}\right\rangle }{2dt}\label{eq4}
\end{equation}
where the set $\left\{ R\right\} $ of walker displacements, each
occurring over the time interval $t$, comprises a Gaussian distribution
that must necessarily be centered well beyond $\xi$. (For practical
purposes, the correlation length $\xi$ is the length scale above
which the \textquotedblleft effective\textquotedblright , or macroscopic,
value of a transport property is obtained.)

{[}It should be clear that the WDM as described here is a mathematical
method\textemdash \textit{not} a model of a physical process. To this
point, the local transport coefficients, which in this paper are local
conductivity values $\sigma_{i}$, may be local values of fluid permeability
$k$ or thermal conductivity $\kappa$, for example.{]}

For displacements $R<\xi$, the walker diffusion is anomalous rather
than Gaussian due to the heterogeneity of the composite at length
scales less than $\xi$. There is, however, an additional characteristic
length $\xi_{0}<\xi$ below which the composite is effectively homogeneous
\citep{VS99a}; this may correspond, for example, to the average phase
domain size. A walker displacement of $\xi$ requiring a travel time
$t_{\xi}=\xi^{2}/(2dD_{w})$ is then comprised of $(\xi/\xi_{0})^{d_{w}}$
segments of length $\xi_{0}$, each requiring a travel time of $t_{0}=\xi_{0}^{2}/(2dD_{0})$,
where $D_{0}$ is the walker diffusion coefficient calculated from
displacements $R\leq\xi_{0}$. Setting $t_{\xi}=(\xi/\xi_{0})^{d_{w}}\,t_{0}$
gives the relation
\begin{equation}
D_{w}=D_{0}\left(\frac{\xi}{\xi_{0}}\right)^{2-d_{w}}=\left(\frac{\xi_{0}^{d_{w}}}{2dt_{0}}\right)\xi^{2-d_{w}}\label{eq5}
\end{equation}
between the walker diffusion coefficient $D_{w}$ (for walks of displacement
$R\geq\xi$), the fractal dimension $d_{w}$ of the walker path (for
walks of displacement $\xi_{0}<R<\xi$), and the correlation length
$\xi$.

\section{Two-component percolation problem}

From the point of view of the WDM, the two-component percolation problem
differs from the conductor/insulator percolation problem mainly by
the fact that walkers are never \textquotedblleft stranded\textquotedblright{}
on finite clusters of conductor sites (until precisely $r=0$). Thus
the approach to the endpoint, which in both cases is percolation only
via the incipient infinite cluster, reflects that difference and so
produces a different set of critical exponents.

Combining Eqs. (\ref{eq1}), (\ref{eq3}) and (\ref{eq5}) gives the
relation
\begin{equation}
r^{u}=\frac{\left\langle \sigma\right\rangle }{\sigma_{1}}\left(\frac{\xi_{0}^{d_{w}}}{2dt_{0}}\right)\xi^{2-d_{w}}\label{eq6}
\end{equation}
which upon rearrangement produces
\begin{equation}
\xi=\left(\frac{\left\langle \sigma\right\rangle }{\sigma_{1}}\right)^{-1/(2-d_{w})}\left(\frac{\xi_{0}^{d_{w}}}{2dt_{0}}\right)^{-1/(2-d_{w})}r^{u/(2-d_{w})}.\label{eq7}
\end{equation}
Thus the correlation length $\xi$ diverges as
\begin{equation}
\xi\sim r^{u/(2-d_{w}^{\text{\dag}})}\label{eq8}
\end{equation}
near $r=0$. The exponent $d_{w}^{\text{\dag}}$ is the limit of the
walker path dimension $d_{w}$ at $r=0$. Surprisingly, it appears
again in the presentation of the conductor/insulator percolation problem,
where its numerical value can be ascertained.

A constraint on the value of the conductivity exponent $u$ arises
from the fact that walkers move according to rules based on \textit{ratios}
of conductivities, and thus $D_{w}$ is a function of those ratios.
This is embodied in the relationship
\begin{multline}
\sigma=\left\langle \sigma(\mathbf{r})\right\rangle D_{w}=\sigma_{1}\left[p_{c}+\frac{\sigma_{2}}{\sigma_{1}}(1-p_{c})\right]D_{w}\\
=\sigma_{2}\left[\frac{\sigma_{1}}{\sigma_{2}}p_{c}+(1-p_{c})\right]D_{w}\label{eq9}
\end{multline}
which simplifies to 
\begin{equation}
\sigma(p_{c},1;(1-p_{c}),r)=r\:\sigma(p_{c},r^{-1};(1-p_{c}),1).\label{eq10}
\end{equation}
Note that the conductivity $\sigma$ on the right-hand side of this
equation diverges as $r\rightarrow0$. Thus
\begin{equation}
\sigma(p_{c},r^{-1};(1-p_{c}),1)=r^{u-1}\label{eq11}
\end{equation}
where the exponent $u-1$ is necessarily less than zero for all dimensions
$d$. In fact this result proves $u_{d}<1$.

The exact value of exponent $u_{2}$ is obtained in the following
way. Note that two random, isotropic systems $(p,\alpha;q,\beta)$
and $(p,\alpha^{-1};q,\beta^{-1})^{\dagger}$ {[}the presence or absence
of the dagger identifies the system{]} are \textit{dual} if the conductivity
of one equals the resistivity of the other. The 2D square bond network,
which has the percolation threshold $p_{c}=\nicefrac{1}{2}$, is known
to be self-dual \citep{Straley}; thus
\begin{equation}
\sigma(\nicefrac{1}{2},1;\nicefrac{1}{2},r)\:\sigma(\nicefrac{1}{2},1;\nicefrac{1}{2},r^{-1})=1.\label{eq12}
\end{equation}
Then
\begin{equation}
\sigma(\nicefrac{1}{2},1;\nicefrac{1}{2},r)\:\sigma(\nicefrac{1}{2},r;\nicefrac{1}{2},1)=r\label{eq13}
\end{equation}
which shows that $\sigma(\nicefrac{1}{2},1;\nicefrac{1}{2},r)=r^{1/2}$,
meaning $u=\nicefrac{1}{2}$. Due to universality, the bond and site
implementations of the two-component percolation problem possess the
same set of critical exponents $\{u_{d}\}$, so $u_{2}=\nicefrac{1}{2}$.

\begin{figure}
\includegraphics[scale=0.55]{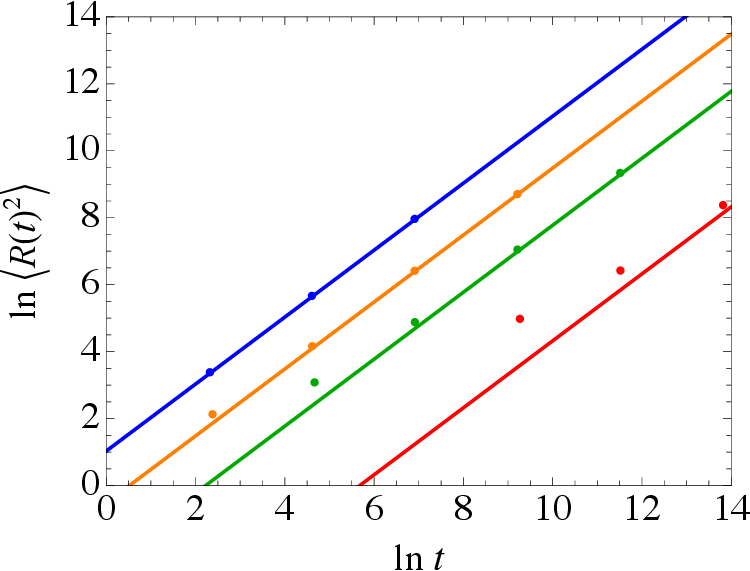}

\caption{\label{fig1}Data supporting the conjecture that the 3D conductivity
exponent $u_{3}$ for the two-component percolation problem equals
$\nicefrac{3}{4}$. The values $\left\langle R(t)^{2}\right\rangle $,
each an average over a different set of $10^{5}$ two-component systems,
are obtained by the WDM; the points would lie on the straight lines
(corresponding to $r=0.1,10^{-2},10^{-3},10^{-5}$, in order from
left to right) in the event that $u_{3}=\nicefrac{3}{4}$. The points
that lie above the straight lines are obtained from walks of displacement
$R(t)<\xi$ and so do not figure in the determination of the value
$u_{3}$.}

\end{figure}

A numerical value for the exponent $u_{3}$ was obtained by the WDM
(details of this sort of calculation are given in Sec. V). Figure
\ref{fig1} shows calculated points $(\ln t,\ln\left\langle R(t)^{2}\right\rangle )$
for two-component systems with $r=0.1$, $10^{-2}$, $10^{-3}$, $10^{-5}$.
The four straight lines of slope $1$ represent the relation $D_{w}=\sigma/\left\langle \sigma\right\rangle $
and so correspond to equations
\begin{equation}
y=x+\ln[2\,d\,D_{w}(r)]=x+\ln\left[\frac{2\,d\,r^{u}}{p_{c}+(1-p_{c})\,r}\right]\label{eq14}
\end{equation}
for the four values of $r$, with the dimension $d$ set to $3$ and
the exponent $u_{3}$ set to the value $0.75$. The coincidence of
the points and the lines support a previous conjecture \citep{VS03}
that $u_{3}=\nicefrac{3}{4}$.

Note, in Fig. \ref{fig1}, that the correlation length $\xi(r)$ increases
as the parameter $r$ declines toward its critical value ($r=0$),
as characteristic of critical phenomena.

The analytical results $u_{d}<1$ and $u_{2}=\nicefrac{1}{2}$ together
with the conjectured result $u_{3}=\nicefrac{3}{4}$ suggest the relations
$u_{d+1}=(u_{d}+1)/2$ and
\begin{equation}
u_{d}=1-(1-u_{2})^{d-1}\label{eq15}
\end{equation}
between the conductivity exponents of the two-component percolation
problem.

\section{Conductor/insulator percolation problem}

The conductor/insulator system has effective conductivity $\sigma=\sigma_{1\,}p\,D_{w}$
where $p$ is the fraction of conductor sites. As the walker diffusion
coefficient $D_{w}=\left\langle R(t)^{2}\right\rangle /(2dt)$ with
walk time $t\gg t_{\xi}$ is obtained from walkers on all conductor
sites, not just those on the percolating cluster, the conductor/insulator
problem is recast as a two-component problem. Namely, the insulator
sites become conductor sites with very low conductivity value $\sigma_{2}\ll\sigma_{1}$.
Then the conductivity exponent $t$ is obtained in the limit $\sigma_{2}=0$
(that is, $r=0$) at $p=p_{c}$. Thus the correlation length for this
two-component system is
\begin{eqnarray}
\xi & = & \left(\frac{\xi_{0}^{d_{w}}}{2dt_{0}}\right)^{-1/(2-d_{w})}D_{w}^{1/(2-d_{w})}\nonumber \\
 &  & \sim p^{-1/(2-d_{w}^{\text{\dag}})}(p-p_{c})^{t/(2-d_{w}^{\text{\dag}})}.\label{eq16}
\end{eqnarray}

It is also the case that $\sigma=\sigma_{1}\,p'\,D'_{w}$ where $p'$
is the fraction of system sites comprising the percolating cluster,
and $D'_{w}$ is the diffusion coefficient for walkers on the percolating
cluster. When the system is near the critical point, $p'$ exhibits
the critical behavior $p'\sim(p-p_{c})^{\beta}$. Thus
\begin{eqnarray}
\xi & \sim & (p-p_{c})^{-\beta/(2-d_{w}^{*})}(p-p_{c})^{t/(2-d_{w}^{*})}\nonumber \\
 &  & \sim(p-p_{c})^{-\nu}.\label{eq17}
\end{eqnarray}
Here the exponent relation $-\nu=(t-\beta)/(2-d_{w}^{*})$ is obtained,
where
\begin{equation}
d_{w}^{*}=2+(t-\beta)/\nu\label{eq18}
\end{equation}
is the limit of the walker path dimension $d_{w}$ at $p=p_{c}$.
(Thus $d_{w}^{*}$ is the fractal dimension of the walker path on
the incipient infinite cluster.) Note that the walker path dimensions
$d_{w}^{*}$ and $d_{w}^{\text{\dag}}$ are related by $d_{w}^{\text{\dag}}-d_{w}^{*}=\beta/\nu$,
and that $d_{w}^{\text{\dag}}=2+t/\nu$.

{[}A more succinct derivation of the exponent relation Eq. (\ref{eq18})
is $\sigma(\xi)=\sigma_{1}\,p'(\xi)\,D'_{w}(\xi)$ implies $\xi^{-t/\nu}\sim\xi^{-\beta/\nu}\,\xi^{2-d_{w}^{*}}$.{]}

The exponents pertaining to the incipient infinite cluster are additionally
connected by a hyperscaling law (a relation that includes the dimension
$d$ of the system). This follows from the asymptotic relation $p'\sim\xi^{-\beta/\nu}$
and the observation that at $p=p_{c}$,
\begin{equation}
p'=\frac{\xi^{D}}{\xi^{d}}\label{eq19}
\end{equation}
where the right-hand side is the volume fraction occupied by the incipient
infinite cluster, the exponent $D$ being the fractal \textquotedblleft mass
dimension\textquotedblright{} of that cluster. Thus
\begin{equation}
\beta=-\nu(D-d).\label{eq:19a}
\end{equation}

The appearance of the critical exponent $d_{w}^{\text{\dag}}$ in
both the two-component percolation problem and the conductor/insulator
percolation problem points to a fundamental connection between the
two systems. Very near the percolation threshold, the effective conductivity
of the conductor/insulator system exhibits the asymptotic behavior
\begin{equation}
\sigma(p>p_{c},1;(1-p),0)\sim(p-p_{c})^{t}\sim\xi^{-t/\nu}\label{eq20}
\end{equation}
while the effective conductivity of the conductor/superconductor system
exhibits the asymptotic behavior
\begin{equation}
\sigma(p<p_{c},\infty;(1-p),1)\sim\left|p-p_{c}\right|^{-s}\sim\xi^{s/\nu}.\label{eq21}
\end{equation}
The exponents $t$ and $s$ can be related to $u$ and $u-1$ from
the two-component percolation problem by noting that the conductivities
of the two conducting systems $\mbox{\ensuremath{(p>p_{c},1;(1\text{\textminus}p),0)}}$
and $(p_{c},1;(1-p_{c}),r)$ are identical, and the conductivities
of the two superconducting systems $(p<p_{c},\infty;(1-p),1)$ and
$(p_{c},r^{-1};(1-p_{c}),1)$ are identical, when the parameters $p$
and $r$ are very close to $p_{c}$ and $0$, respectively. That is,
\begin{equation}
\sigma(p_{c},1;(1-p_{c}),r)\sim\sigma(p>p_{c},1;(1\text{\textminus}p),0)\sim\xi^{-t/\nu}\label{eq23}
\end{equation}
and
\begin{equation}
\sigma(p_{c},r^{-1};(1-p_{c}),1)\sim\sigma(p<p_{c},\infty;(1\text{\textminus}p),1)\sim\xi^{s/\nu}.\label{eq25}
\end{equation}
Note that Eq. (\ref{eq10}) produces the relation
\begin{equation}
\sigma(p_{c},1;(1-p_{c}),r)=\left[\sigma(p_{c},r^{-1};(1-p_{c}),1)\right]^{u/(u-1)}\label{eq26}
\end{equation}
which has the asymptotic expression 
\begin{equation}
\xi^{-t/\nu}=\left[\xi^{s/\nu}\right]^{u/(u-1)}\label{eq26a}
\end{equation}
thereby revealing the exponent relation
\begin{equation}
\frac{t}{s}=\frac{u}{1-u}\label{eq27}
\end{equation}
or equivalently $u=t/(s+t)$, in all dimensions.

In fact the four conductivities in Eqs. (\ref{eq23}) and (\ref{eq25})
have in common the asymptotic relationship $D_{w}\sim\xi^{2-d_{w}^{\text{\dag}}}$.
This is because very near $r=0$ and $p=p_{c}$ the four systems have
(statistically) identical morphologies, and identical phase conductivity
ratios. Thus a diffusing walker finds the four systems identical.
For example, the effective conductivity
\begin{align}
\sigma(p & <p_{c},\infty;(1\text{\textminus}p),1)\sim\sigma(p<p_{c},r^{-1};(1\text{\textminus}p),1)\nonumber \\
 & \sim r^{-1}\,\xi^{2-d_{w}^{\text{\dag}}}\sim\xi^{(t/\nu)/u}\,\xi^{-t/\nu}=\xi^{s/\nu}.\label{eq24}
\end{align}
Similarly, the conductivities in Eqs. (\ref{eq20}) and (\ref{eq21})
have in common the asymptotic relationship $D'_{w}\sim\xi^{2-d_{w}^{*}}$.

Given the exponent relation Eq. (\ref{eq27}), a consequence of Eq.
(\ref{eq15}) is
\begin{equation}
\frac{t_{d}}{s_{d}}=2^{d-1}-1.\label{eq28}
\end{equation}
Using the value for exponent $t_{3}$ calculated in the following
section, the value $s_{3}=0.67787(105)$ is a prediction.

It is interesting to consider a counterpart to Eq. (\ref{eq15}) for
the conductor/insulator system. In this case the conductivity exponent
$t_{d}$ increases towards $3$ as the dimension increases \citep{SA}.
Then
\begin{equation}
t_{d}=3\left[1-\left(1-\frac{t_{2}}{3}\right)^{d-1}\right].\label{eq29}
\end{equation}
Given the generally accepted value $t_{2}=1.30$ ($1.299$), this
equation produces $t_{3}=2.03667$ ($2.03553$) and similarly reasonable
values for higher dimensions.

\section{Numerical approach and results}

Because the critical exponents are obtained from the incipient infinite
cluster, it is important to ensure that the diffusing walkers, which
perform the calculations, are indeed on that cluster. To start, a
walker is placed on a conductor site at the center of a vast volume
of \textquotedblleft undefined\textquotedblright{} sites. Then each
neighboring site is defined to be conducting (with probability $p_{c}$)
or is otherwise insulating. Rather than have the walker then attempt
a move to a randomly chosen neighboring site (which may not be successful),
it is more efficient to utilize the variable residence time algorithm,
which takes advantage of the statistical nature of the diffusion process.

According to this algorithm \citep{VS99}, the actual behavior of
the walker is well approximated by a sequence of moves in which the
direction of the move from a site $i$ is determined randomly by the
set of probabilities $\left\{ P_{i\rightarrow j}\right\} $, where
$P_{i\rightarrow j}$ is the probability that the move is to adjacent
site $j$ (which has conductivity $\sigma_{j}$) and is given by the
equation
\begin{equation}
P_{i\rightarrow j}=\frac{\sigma_{j}}{\sigma_{i}+\sigma_{j}}\left[\sum_{k=1}^{2d}\left(\frac{\sigma_{k}}{\sigma_{i}+\sigma_{k}}\right)\right]^{-1}.
\end{equation}
The sum is over all sites adjacent to site $i$. The time interval
over which the move occurs is
\begin{equation}
T_{i}=\left[2\sum_{k=1}^{2d}\left(\frac{\sigma_{k}}{\sigma_{i}+\sigma_{k}}\right)\right]^{-1}.\label{eq32}
\end{equation}
Note that this version of the variable residence time algorithm is
intended for orthogonal systems (meaning a site in a 3D system has
six neighbors, for example).

After each move, any \textquotedblleft undefined\textquotedblright{}
neighboring sites are converted to conducting or insulating. In this
way the cluster grows. A walk is complete when the sum of move times
$T_{i}$ reaches or exceeds a preset walk time $T$.

Of course, many of those clusters turn out to be finite and so clearly
are not part of the incipient infinite cluster. Indeed, the larger
the preset walk time $T$, the greater the likelihood that a nascent
cluster will turn out to be finite. Note that finite and infinite
clusters are easily distinguished: A finite cluster has the characteristic
that all conductor sites comprising the cluster have been visited
by time $T$, indicating the cluster is completely surrounded by insulator
sites. An ``infinite'' (or percolating) cluster includes at least
one conductor site that was \textquotedblleft created\textquotedblright{}
by the walker (in the manner described above) but never actually visited
in time $T$.

In general, $n\times10^{5}$ \textquotedblleft infinite\textquotedblright{}
clusters for each walk time $T$ were used to determine the value
of a critical exponent or a ratio of exponents. These represent $n\times10^{5}$
\textit{different pieces}, each of size corresponding to the walk
time $T$, of the incipient infinite cluster. It doesn\textquoteright t
matter that a cluster still \textquotedblleft infinite\textquotedblright{}
at time $T$ might turn out to be finite were the walk extended to
longer times, since every finite cluster at the percolation threshold
resembles the incipient infinite cluster (which is statistically self-similar
over \textit{all} length scales) over length scales up to the size
of the cluster.

The numerical data recorded for the incipient infinite cluster was,
for each of several preset walk times $T$, the following: (1) The
number $N_{\mathrm{pc}}=10^{5}$ of percolating (\textquotedblleft infinite\textquotedblright )
clusters over which most other quantities are averaged. (2) The number
$N_{\mathrm{fc}}$ of finite clusters encountered in the process of
accumulating $N_{\mathrm{pc}}$ percolating clusters. (3) The actual
(averaged) walk time $t$ (very slightly larger than $T$). (4) The
average walker displacement $\left\langle R(t)\right\rangle $. (5)
The average walker displacement-squared $\left\langle R(t)^{2}\right\rangle $.
(6) The average number $\left\langle n_{m}(t)\right\rangle $ of walker
moves. (7) The average number $\left\langle n_{s}(t)\right\rangle =\left\langle S(t)\right\rangle $
of visited sites.

The percolation threshold values used in the calculations are $p_{c}=0.592746$
(2D) and $p_{c}=0.311607$ (3D). The \textquotedblleft standard\textquotedblright{}
values for $\beta$, $\nu$, and $D$ referred to below are $\beta_{2}=\nicefrac{5}{36}$,
$\nu_{2}=\nicefrac{4}{3}$, $D_{2}=\nicefrac{91}{48}$ \citep{SA};
and $\beta_{3}=0.41810(57)$, $\nu_{3}=0.87642(115)$, $D_{3}=2.52295(15)$,
derived from values $1/\nu_{3}=1.1410(15)$ and $\beta_{3}/\nu_{3}=0.47705(15)$
\citep{Wang}.

\subsection{Comment on average value $\left\langle R(t)^{2}\right\rangle $}

\begin{figure}
\includegraphics[scale=0.55]{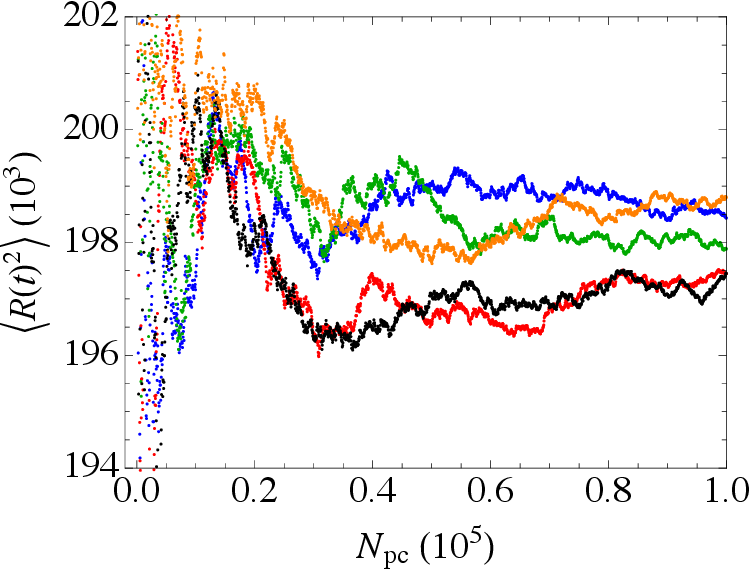}

\caption{\label{fig2}Sets of points (distinguished by color) that converge
toward a \textquotedblleft correct\textquotedblright{} value for the
average walker displacement-squared $\left\langle R(t)^{2}\right\rangle $
for walks of time $t=10^{7}$ over the incipient infinite cluster
in 2D. The variable $N_{\mathrm{pc}}$ corresponds to the number of
independent walks from which the average value $\left\langle R(t)^{2}\right\rangle $
is obtained. }
\end{figure}

\begin{figure}
\includegraphics[scale=0.55]{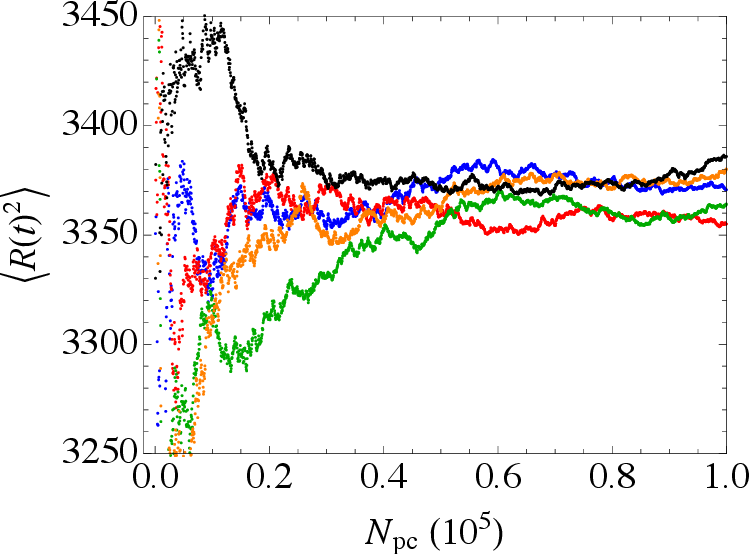}

\caption{\label{fig3}Sets of points (distinguished by color) that converge
toward a \textquotedblleft correct\textquotedblright{} value for the
average walker displacement-squared $\left\langle R(t)^{2}\right\rangle $
for walks of time $t=10^{6}$ over the incipient infinite cluster
in 3D. The variable $N_{\mathrm{pc}}$ corresponds to the number of
independent walks from which the average value $\left\langle R(t)^{2}\right\rangle $
is obtained. }

\end{figure}

Most calculations of interest require arguably correct (as well as
accurate) values for the average walker displacement-squared $\left\langle R(t)^{2}\right\rangle $.
In particular it is important that a sufficient number of independent
walks (i.e., walks over a sufficient number of distinct sections of
a percolating system) be taken in order that a mean value for $\left\langle R(t)^{2}\right\rangle $
with reasonably narrow bounds is obtained. Figures \ref{fig2} and
\ref{fig3} are instructive on this point.

Figure \ref{fig2} shows five sets of points (distinguished by color)
pertaining to walker diffusion on the incipient infinite cluster in
2D. Consider \textit{one} of those sets: The coordinates of the points
are $(N_{\mathrm{pc}},\left\langle R(t)^{2}\right\rangle )$, where
the average value $\left\langle R(t)^{2}\right\rangle $ is obtained
from $N_{\mathrm{pc}}$ percolating clusters (that is, from $N_{\mathrm{pc}}$
independent walks). As more walks are taken (i.e., as $N_{\mathrm{pc}}$
increases), the average value $\left\langle R(t)^{2}\right\rangle $
fluctuates less and flattens out. Then by creating several sets and
reproducing this behavior, a set size $N_{\mathrm{pc}}$ is found
($10^{5}$ in this case) that permits a mean value $\left\langle \left\langle R(t)^{2}\right\rangle \right\rangle $
to be obtained with reasonably narrow bounds.

Similarly, Fig. \ref{fig3} shows five sets of points pertaining to
walker diffusion on the incipient infinite cluster in 3D. Again, sets
of size $N_{\mathrm{pc}}=10^{5}$ appear to be sufficient to obtain
a defensible value for $\left\langle R(t)^{2}\right\rangle $ for
use in calculations. (Larger sets may naturally reduce the bounds,
but at the cost of significantly increased computer time.)

Data from Figs. \ref{fig2} (walk time $t=10^{7}$) and \ref{fig3}
($t=10^{6}$) are used (together with additional sets of size $10^{5}$)
in the calculations of $d_{w}^{*}$ below.

\subsection{Walker path dimension $d_{w}^{*}$}

For percolating systems of size $L<\xi$, the equivalent of Eq. (\ref{eq5})
is
\begin{equation}
D_{w}(L)=D_{w}(\xi)\left(\frac{L}{\xi}\right)^{2-d_{w}}=\left(\frac{\xi_{0}^{d_{w}}}{2dt_{0}}\right)L^{2-d_{w}}.\label{eq33}
\end{equation}
In the case of the incipient infinite cluster, which is statistically
self-similar over all length scales, this relation can be expressed
in terms of the computable variable $\left\langle R(t)^{2}\right\rangle $,
namely,
\begin{eqnarray}
\frac{\left\langle R(t)^{2}\right\rangle }{2dt} & = & \left(\frac{\xi_{0}^{d_{w}}}{2dt_{0}}\right)\left\langle R(t)^{2}\right\rangle ^{1-d_{w}^{*}/2}\nonumber \\
 &  & =\left\langle R(t)^{2}\right\rangle ^{1-d_{w}^{*}/2}.\label{eq34}
\end{eqnarray}
The last equality comes about because the characteristic length $\xi_{0}$
is the size of a single conductor site; that is, $\xi_{0}=1$. This
Gaussian regime corresponds to walkers diffusing within the conductor
site for walk times $t<t_{0}$. Then the diffusion coefficient $D_{0}=1$
and so the travel time $t_{0}=(2d)^{-1}$. Thus
\begin{equation}
\left\langle R(t)^{2}\right\rangle =(2dt)^{2/d_{w}^{*}}\label{eq35}
\end{equation}
or equivalently,
\begin{equation}
\ln\left\langle R(t)^{2}\right\rangle =\frac{2}{d_{w}^{*}}\ln t+\frac{2}{d_{w}^{*}}\ln(2d).\label{eq36}
\end{equation}

\begin{figure}
\includegraphics[scale=0.55]{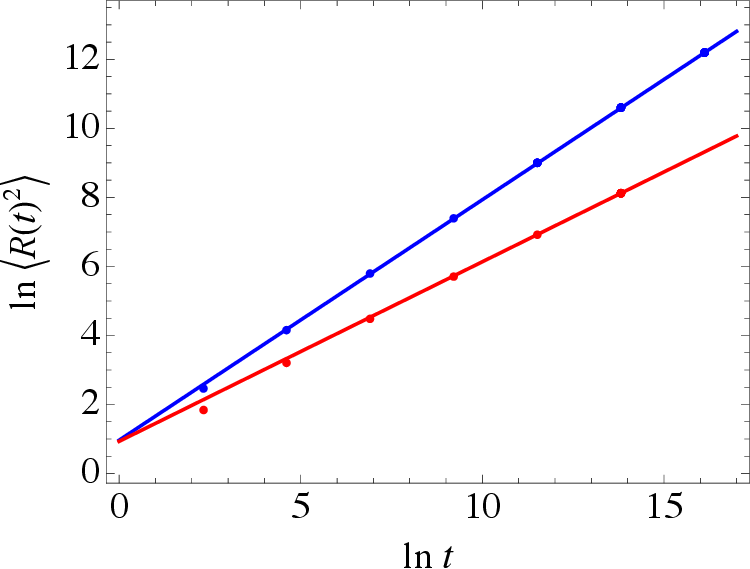}

\caption{\label{fig4}Data obtained from walks over the 2D and 3D incipient
infinite clusters by which the walker path dimension $d_{w}^{*}$
is calculated. Both straight lines have slope $2/d_{w}^{*}$; the
line of greater (lesser) slope corresponds to the 2D (3D) incipient
infinite cluster. The value $\left\langle R(t)^{2}\right\rangle $
for each point is obtained from one or more sets of $10^{5}$ distinct
sections of the incipient infinite cluster. Points at short walk times
$t$ are affected by the finite size of the conductor sites.}

\end{figure}

This last equation produces the straight lines in Fig. \ref{fig4}.
The line of greater (lesser) slope, running through the point corresponding
to largest walk time $t$, has slope inversely proportional to the
walker path dimension $d_{w}^{*}$ for 2D (3D) percolation. Note that
in both cases, points corresponding to shorter walk times lie below
the straight lines, due to the effect of the finite (not infinitesimal)
size of the conductor sites. {[}A more precise explanation is as follows:
Walker diffusion on the incipient infinite cluster comprised of conductor
sites is Gaussian ($d_{w}=2$) for walk times $t<t_{0}$, and anomalous
($d_{w}=d_{w}^{*}>2$) for walk times $t>t_{0}$, so that lines of
slope $1$ and slope $2/d_{w}^{*}$ meet at the point $(\ln t_{0},\ln\left\langle R(t_{0})^{2}\right\rangle )=(-\ln2d,\ln1)$.
Points in the anomalous regime near $t=t_{0}$ are thus affected by
the presence of the Gaussian regime and so lie \textit{below} the
slope $2/d_{w}^{*}$ line.{]}

For 2D percolation, the value $d_{w}^{*}=2.87038(60)$ was obtained
from $10^{6}$ walks, each of duration $T=10^{7}$, over ten sets
of $10^{5}$ clusters (representing $10^{6}$ distinct sections of
the incipient infinite cluster). The average number of moves per walk
$\left\langle n_{m}\right\rangle >25\times10^{6}$, and the average
number of visited sites per walk $\left\langle n_{s}\right\rangle >68\times10^{3}$.

For 3D percolation, the value $d_{w}^{*}=3.84331(193)$ was obtained
from $8\times10^{5}$ walks, each of duration $T=10^{6}$, over eight
sets of $10^{5}$ clusters (representing $8\times10^{5}$ distinct
sections of the incipient infinite cluster). The average number of
moves per walk $\left\langle n_{m}\right\rangle >2.3\times10^{6}$,
and the average number of visited sites per walk $\left\langle n_{s}\right\rangle >12\times10^{3}$.

In both cases Fig. \ref{fig4} shows that these walks are of sufficient
length (sufficient walk time) that finite-site-size effects on these
$d_{w}^{*}$ values are negligible, and Figs. \ref{fig2} and \ref{fig3}
show that a sufficient number of randomly selected sections of the
incipient cluster are explored to give exponent values within meaningful
brackets.

Table \ref{table1} presents values of critical exponents calculated
from these WDM values for $d_{w}^{*}$.

Note that the data from these $n\times10^{5}$ walks over time $T=10^{7}$
(2D) or $10^{6}$ (3D) are used in all the following calculations
that pertain to the incipient infinite cluster. Data for shorter walk
times $T=10,10^{2},10^{3},\ldots$ are obtained from one or more sets
of $10^{5}$ walks.

\begin{table}
\caption{\label{table1}Calculated values for the walker path dimension $d_{w}^{*}$,
the exponent ratio $t/\nu$, the conductivity exponent $t$, and the
spectral dimension $d_{s}=2D/d_{w}^{*}$.}

\begin{tabular}{ccccc}
\toprule 
 & $d_{w}^{*}$ & $t/\nu$ & $t$ & $d_{s}$\tabularnewline
\midrule
\midrule 
2D & 2.87038(60) & 0.974542(600) & 1.29939(80) & 1.32097(28)\tabularnewline
\midrule
\midrule 
3D & 3.84331(193) & 2.32036(193) & 2.0336(32) & 1.3129(7)\tabularnewline
\bottomrule
\end{tabular}
\end{table}

\subsection{Incipient infinite cluster mass dimension $D$}

A lower bound $D_{s}$ on the mass dimension $D$ of the incipient
infinite cluster is found by considering the number $S(t)$ of distinct
sites visited during a walk to be proportional to $R_{c}(t)^{D_{s}}$,
where $R_{c}(t)$ is the crude radius of the cluster of visited sites.
This cluster radius can be related to the walker displacement $R(t)$
by noting that the walker is essentially equilibrated after many moves
over the cluster of visited sites ($n_{m}/n_{s}\gg1$). Then the displacement
$R(t)$ finds the walker at any site of the cluster with equal probability.
For example, in the case of a walker confined to a 3D spherical cluster
of conductor sites, the average value $\left\langle r\right\rangle $
is given by
\begin{equation}
\left\langle r\right\rangle =\left(\frac{4}{3}\pi R_{c}^{3}\right)^{-1}\int_{r=0}^{R_{c}}r\cdot4\pi r^{2}dr=\frac{3}{4}R_{c}\label{eq37}
\end{equation}
since $r$, that is $R(t)$, is measured from the origin of the cluster
(the original site from which the cluster grew). More generally, $R_{c}\propto\left\langle R(t)\right\rangle $
and therefore
\begin{equation}
\left\langle S(t)\right\rangle \propto\left\langle R(t)\right\rangle ^{D_{s}}\label{eq38}
\end{equation}
with the averages obtained from a very large number of clusters and
walks.

\begin{figure}
\includegraphics[scale=0.55]{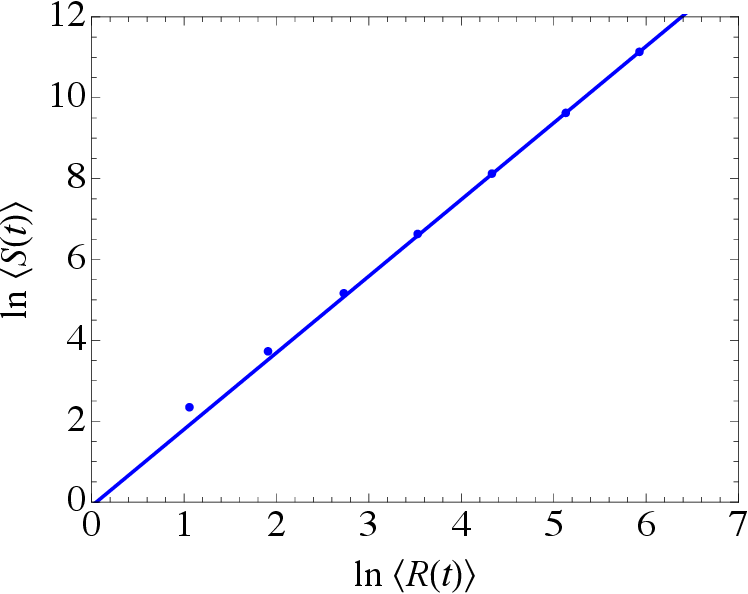}

\caption{\label{fig5}Data obtained from walks over the 2D incipient infinite
cluster by which the fractal dimension $D_{s}$ of the cluster $S(t)$
of visited sites is calculated. The straight line fit to points for
$t=10^{6}$ and $10^{7}$ has slope $D_{s}$, giving a lower bound
for the fractal dimension $D$ of the incipient infinite cluster.
The values $\left\langle R(t)\right\rangle $ and $\left\langle S(t)\right\rangle $
for each point are obtained from one or more different sets of $10^{5}$
distinct sections of the incipient infinite cluster. Points at short
walk times $t$ are affected by the finite size of the conductor sites.}

\end{figure}

\begin{figure}
\includegraphics[scale=0.55]{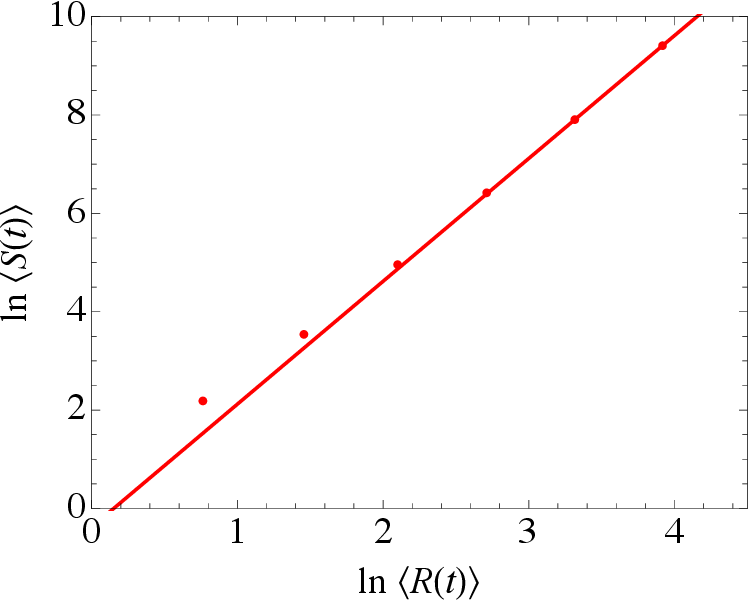}

\caption{\label{fig6}Data obtained from walks over the 3D incipient infinite
cluster by which the fractal dimension $D_{s}$ of the cluster $S(t)$
of visited sites is calculated. The straight line fit to points for
$t=10^{5}$ and $10^{6}$ has slope $D_{s}$, giving a lower bound
for the fractal dimension $D$ of the incipient infinite cluster.
The values $\left\langle R(t)\right\rangle $ and $\left\langle S(t)\right\rangle $
for each point are obtained from one or more different sets of $10^{5}$
distinct sections of the incipient infinite cluster. Points at short
walk times $t$ are affected by the finite size of the conductor sites.}

\end{figure}

This relation produces the straight lines in Figs. \ref{fig5} and
\ref{fig6} which describe the growth of the cluster of visited sites
produced by walkers confined to the incipient infinite cluster. In
Fig. \ref{fig5} the slope $D_{s}=1.89503$ is obtained for 2D percolation;
this $D_{s}$ value is slightly less than the fractal dimension $D=91/48=1.89583$
of the incipient infinite cluster \citep{SA}. In Fig. \ref{fig6}
the slope $D_{s}=2.49848$ is obtained for 3D percolation; similarly,
this $D_{s}$ value is slightly less than the standard value $D=2.52295(15)$
for the incipient infinite cluster \citep{Wang}. In both cases the
line was fit to the two largest-walk-time points (each point obtained
from eight or more sets of $10^{5}$ independent walks) in order to
minimize the effects of the finite (not infinitesimal) size of the
conductor sites apparent at shorter times $t$.

While the value $D_{s}$ may be very close to $D$, it will always
be smaller since the cluster $S(t)$ will never completely fill the
section of the incipient infinite cluster explored by the walker over
time $t$ (the walker will never visit every accessible site in that
section). An extreme example of this effect is walker diffusion over
a homogeneous 2D system: the path dimension $d_{w}$ is (correctly)
found to be precisely $2$, but $D_{s}\approx1.885$ (far less than
$D=d=2$) since the cluster $S(t)$ in that case grows in a non-compact
way and so suggests a system with dimension less than $2$.

Note that a variation on Eq. (\ref{eq38}) is
\begin{equation}
S(t)\propto\left\langle R(t)^{2}\right\rangle ^{D/2}=\left[(2dt)^{2/d_{w}^{*}}\right]^{D/2}\propto t^{d_{s}/2}\label{eq38a}
\end{equation}
where the equality is obtained from Eq. (\ref{eq35}), and the spectral
dimension $d_{s}=2D/d_{w}^{*}$. However, this approach is discouraged
as $\left\langle R(t)^{2}\right\rangle ^{1/2}$ is a poor approximation
of $\left\langle R(t)\right\rangle $.

\subsection{Fraction $p'$}

The fraction $p'$ of system sites that comprise the percolating cluster
appears in the expression for conductivity $\sigma=\sigma_{1}\,p'\,D'_{w}$
for systems with $p>p_{c}$, and in the relation $p'\sim(p-p_{c})^{\beta}$
for infinite systems very close to the percolation threshold. An expression
for $p'$ is derived here, to be used in calculations below.

It is reasonable to assume that a created cluster of size greater
than the correlation length $\xi$ (which occurs when the created
cluster is \textquotedblleft infinite\textquotedblright{} at preset
walk time $T\gg t_{\xi}$) is part of the percolating cluster. A very
large number $N_{\mathrm{pc}}$ of such \textquotedblleft infinite\textquotedblright{}
clusters are needed in the calculation of $D'_{w}$. In the process
of creating these $N_{\mathrm{pc}}$ percolating clusters, a number
$N_{\mathrm{fc}}$ of smaller, \textquotedblleft finite\textquotedblright{}
clusters are generated that cannot be used in the calculation of $D'_{w}$.
Recall that creation of each cluster ($N_{\mathrm{pc}}+N_{\mathrm{fc}}$
in total) begins by designating a ``seed'' conductor site within
a vast volume of ``undefined'' sites. A fraction $p'/p$ of those
``seed'' sites will turn out to belong to a percolating cluster.
Thus $p'/p=N_{\mathrm{pc}}/(N_{\mathrm{pc}}+N_{\mathrm{fc}})$, or
equivalently

\begin{equation}
p'=p\left(1+\frac{N_{\mathrm{fc}}}{N_{\mathrm{pc}}}\right)^{-1}.\label{eq39}
\end{equation}

\subsection{Exponent ratio $\beta/\nu$}

The asymptotic relation $p'\sim\xi^{-\beta/\nu}$ inspires the finite-size
scaling relation $p'(L)\propto L^{-\beta/\nu}$ that gives the fraction
of sites in an arbitrary portion of size $L$ of an infinite system
at $p=p_{c}$, that belong to the cluster that percolates the size
$L$ volume.

An equivalent scaling relation is
\begin{equation}
p'(t)\propto\left\langle R(t)\right\rangle ^{-\beta/\nu}\label{eq:40}
\end{equation}
which pertains to walks of time $t$ over clusters created in the
manner described at the beginning of Sec. V, for infinite systems
at $p=p_{c}$. The $N_{\mathrm{pc}}$ walks that produce the set of
$R(t)$ values also give the value of $p'(t)$ as described in the
previous subsection {[}Eq. (\ref{eq39}) with $p=p_{c}${]}. Note
that the value $\left\langle R(t)\right\rangle $ effectively serves
as the correlation length needed for this use of Eq. (\ref{eq39}).

\begin{figure}
\includegraphics[scale=0.55]{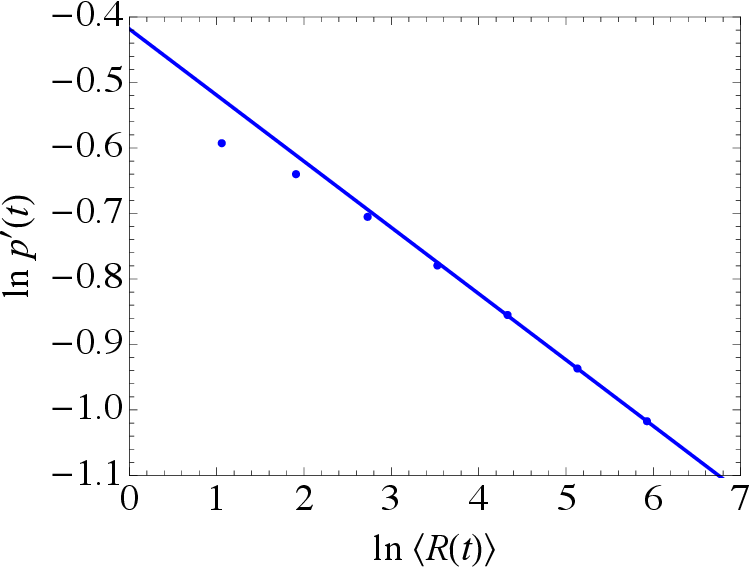}

\caption{\label{fig7}Data obtained from walks over the 2D incipient infinite
cluster by which the exponent ratio $\beta_{2}/\nu_{2}$ is calculated.
Values $\left\langle R(t)\right\rangle $ and $p'(t)$ are obtained
for walk times $t=10,10^{2},\ldots,10^{7}$. The straight line fit
to the two points at the largest walk times has slope approximating
$-\beta_{2}/\nu_{2}$. Points lying below the straight line (at short
walk times) are affected by the finite size of the conductor sites.}

\end{figure}

\begin{figure}
\includegraphics[scale=0.55]{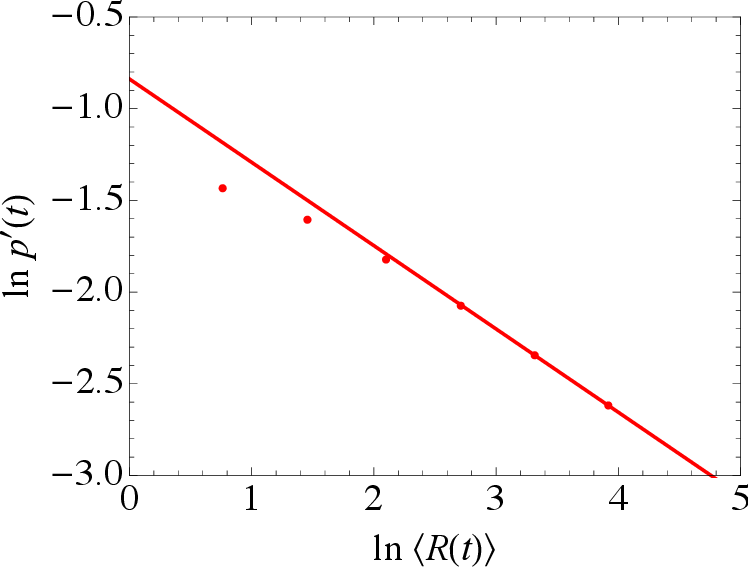}

\caption{\label{fig8}Data obtained from walks over the 3D incipient infinite
cluster by which the exponent ratio $\beta_{3}/\nu_{3}$ is calculated.
Values $\left\langle R(t)\right\rangle $ and $p'(t)$ are obtained
for walk times $t=10,10^{2},\ldots,10^{6}$. The straight line fit
to the two points at the largest walk times has slope approximating
$-\beta_{3}/\nu_{3}$. Points lying below the straight line (at short
walk times) are affected by the finite size of the conductor sites.}

\end{figure}

The novel scaling relation Eq. (\ref{eq:40}) produces the straight
lines (with slope approximating $-\beta/\nu$) in Figs. \ref{fig7}
and \ref{fig8}. In both the 2D (Fig. \ref{fig7}) and 3D (Fig. \ref{fig8})
cases the fits are to the points for the two largest walk times (each
point obtained from eight or more sets of $10^{5}$ independent walks).
These produce values $\beta_{2}/\nu_{2}=0.101027$ (compare to the
exact value $5/48=0.104167$ \citep{SA}) and $\beta_{3}/\nu_{3}=0.454446$
(compare to the value $0.47705(15)$ \citep{Wang}).

A different formulation $p'(t)\propto\left\langle R(t)^{2}\right\rangle ^{-\beta/2\nu}$
gives very similar values: $\beta_{2}/\nu_{2}=0.100952$ and $\beta_{3}/\nu_{3}=0.453645$.

Note that the points for short walk times are affected by the finite
size of the conductor sites. Indeed, in both figures the point $\left(\ln1,\ln p_{c}\right)$
lines up with the plotted points lying below the fitted line.

As points are obtained at ever-larger walk times, the slopes of the
fitted lines will increase in magnitude, giving values for the exponent
ratio $\beta/\nu$ closer to the true ones. This accords with the
\textit{asymptotic} expression of Eq. (\ref{eq:40}),
\begin{equation}
p'(t)\sim\left\langle R(t)\right\rangle ^{-\beta/\nu}.\label{eq:40a}
\end{equation}

\subsection{Conductivity $\sigma$ of percolating systems with $p>p_{c}$}

For this case ($p>p_{c}$) the effective conductivity $\sigma=\sigma_{1}\,p'\,D'_{w}$
where $p'$ is the fraction of system sites comprising the percolating
cluster, and $D'_{w}=\left\langle R(t)^{2}\right\rangle /(2dt)$ is
the diffusion coefficient for walkers on the percolating cluster.
Walk times $t$ should be sufficiently large that $\left\langle R(t)^{2}\right\rangle \gg\xi^{2}$.
{[}Or equivalently, walk times $t$ should be sufficiently large that
$D'_{w}$ has declined to a constant value. Too-small walk times produce
incorrect $D'_{w}$ values, that are too high.{]} The function $p'(t)=p/(1+N_{\mathrm{fc}}/N_{\mathrm{pc}})$
where the ratio $N_{\mathrm{fc}}/N_{\mathrm{pc}}$ is obtained in
the course of generating the large number of walks of time $t\gg t_{\xi}$.

\subsection{Conductivity exponent $u_{3}$}

For the two-component system, the effective conductivity $\sigma=\left\langle \sigma\right\rangle D_{w}$
where $\left\langle \sigma\right\rangle =p_{c}\,\sigma_{1}+(1-p_{c})\,\sigma_{2}$,
and $D_{w}=\left\langle R(t)^{2}\right\rangle /(2dt)$ is the walker
diffusion coefficient obtained for walk times $t\gg t_{\xi}$. While
every system site is accessible to a walker (in contrast to the conductor/insulator
system), it is convenient to use the same \textquotedblleft created
cluster\textquotedblright{} code.

Thus the walker is initially placed on a site that is randomly chosen
to be of the $\sigma_{1}$ sort (with probability $p_{c}$) or is
otherwise the $\sigma_{2}$ sort. Then each neighboring site is defined
to be of the $\sigma_{1}$ sort (with probability $p_{c}$) or is
otherwise the $\sigma_{2}$ sort. Then the walker moves to one of
those sites over a time $T_{i}$ as dictated by the variable residence
time algorithm. And so on.

The 3D results for ratios $\sigma_{2}/\sigma_{1}=0.1,10^{-2},10^{-3},10^{-5}$
are shown in Fig. \ref{fig1}. As discussed near the end of Sec. III,
they support a previous conjecture that $u_{3}=\nicefrac{3}{4}$.

\section{Concluding remarks}

The intent of this research was to clarify the relationship between
the two-component percolation problem and the familiar conductor/insulator
percolation problem. The Walker Diffusion Method provided a new conceptual,
analytical, and numerical approach to this task.

An important achievement is the recognition of a new critical exponent
$d_{w}^{\text{\dag}}$ that connects the two types of percolating
systems. This is the fractal dimension of the walker path in the two-component
system at the endpoint $r=0$. It is also the limit of the walker
path dimension $d_{w}$ in the conductor/insulator system when all
conductor clusters are connected by an extremely low conductivity
\textquotedblleft background\textquotedblright{} (replacing the insulator
phase), attained at $p=p_{c}$ and background conductivity reduced
to zero. The connection made apparent by $d_{w}^{\text{\dag}}$ leads
to Eq. (\ref{eq27}), relating the conductivity exponent $t$ and
superconductivity exponent $s$, and the corresponding exponents $u$
and $1-u$.

The value $d_{w}^{\text{\dag}}$ is best calculated from the exponent
relation $d_{w}^{\text{\dag}}=2+t/\nu$ derived in Sec. IV. Use of
the calculated value for $t_{2}$ and the standard value for $\nu_{2}$
produce $d_{w}^{\text{\dag}}=2.97454(60)$ for 2D systems. In principle
$d_{w}^{\text{\dag}}$ may also be obtained via the relation
\begin{equation}
\left\langle R(t)^{2}\right\rangle =(2dt)^{2/d_{w}^{\text{\dag}}}\label{eq41}
\end{equation}
describing walks over the conductor/insulator system at $p=p_{c}$,
where walkers on the finite clusters (in addition to those on the
incipient infinite cluster) are included in the calculation. Those
trapped walkers diffuse according to the variable residence time algorithm
during the walk time $t$, and so contribute to the average displacement-squared
$\left\langle R(t)^{2}\right\rangle $ (hence $d_{w}^{\text{\dag}}>d_{w}^{*}$).

Additionally, very good values for the critical exponent $d_{w}^{*}$
in two and three dimensions are obtained, which enable calculation
of accurate values for the conductivity exponents $t_{2}$ and $t_{3}$.
WDM calculations also support the conjectured value $u_{3}=\nicefrac{3}{4}$,
which motivates a proposed set of equations connecting conductivity
exponents across dimensions.
\begin{acknowledgments}
I thank Professor Robert ``Bob'' Smith (Department of Geological
Sciences) for arranging my access to the resources of the University
of Idaho Library (Moscow, Idaho).
\end{acknowledgments}

\appendix

\section{WDM for bond-based systems}

In this case \citep{VS02}, the walkers reside on the zero-dimensional
nodes of a regular network of bonds. The principle of detailed balance
ensures that at equilibrium (i.e., no walker sources or sinks) a uniform
walker density $\rho_{i}=1$ is maintained. This is implemented by
a variable residence time algorithm whereby every attempted move from
a node is successful but the move is accomplished over a variable
time interval. Specifically, the direction of each move from a node
$i$ (to a connected node $j$) is determined randomly by the set
of probabilities $\left\{ P_{i\rightarrow j}\right\} $, where
\begin{equation}
P_{i\rightarrow j}=\frac{\sigma_{ij}}{\sum_{k}\sigma_{ik}}\label{A1}
\end{equation}
and the set $\left\{ \sigma_{ik}\right\} $ are the conductivities
of the bonds connecting node $i$ and node $k$. The time interval
over which the move occurs is
\begin{equation}
T_{i}=\frac{\phi}{\sum_{k}\sigma_{ik}}\label{A2}
\end{equation}
where $\phi=1$ in the case of orthogonal networks (e.g., square and
cubic networks) and $\phi=3/2$ in the case of triangular 2D networks.
The paths of the walkers thus reflect the distribution and conductivity
of the conducting bonds, and may be described at the macroscopic scale
by the walker diffusion coefficient $D_{w}$. That is related to the
effective conductivity $\sigma$ by
\begin{equation}
\sigma=f_{w}\,D_{w}\label{A3}
\end{equation}
where the factor $f_{w}$ is the fraction of walkers that are mobile
(so equal to the fraction of nodes that have at least one attached
conductor bond). The value $D_{w}$ is calculated from the equation
\begin{equation}
D_{w}=\frac{\left\langle R(t)^{2}\right\rangle }{2dt}\label{A4}
\end{equation}
where $d$ is the Euclidean dimension of the network; and the set
$\left\{ R\right\} $ of walker displacements, each occurring over
the time interval $t$, comprises a Gaussian distribution that must
necessarily be centered at $\left\langle R\right\rangle \gg\xi$.

Equation (\ref{eq5}) applies to bond-based systems as well. In particular,
\begin{equation}
D_{w}=D_{0}\left(\frac{\xi}{\xi_{0}}\right)^{2-d_{w}}\label{A5}
\end{equation}
where $D_{0}$ is the walker diffusion coefficient calculated from
displacements $R\leq\xi_{0}$. It is evident from Eq. (\ref{A2})
that $D_{0}$ has a conductivity value. For example, walks over the
incipient infinite cluster have $\xi_{0}$ equal to one bond length,
so $D_{0}=\sigma_{1}$.

In order to use Eq. (\ref{A3}) to obtain $\sigma$ for a particular
system, the fraction $f_{w}$ of ``active'' nodes must be ascertained.
Obviously $f_{w}=1$ in the case of the two-component percolation
problem. Another example is the bond-and-node Sierpinski triangle
\citep{SierTri}, where the conductivity properties (critical exponents
and dimensions at the limit of recursion iteration $i\rightarrow\infty$)
are obtained by considering an infinite 2D array of Sierpinski triangles.
In that case an algebraic formula is derived for $f_{w}$ as a function
of iteration number $i$.

The conductor/insulator bond percolation problem is addressed by the
relation
\begin{equation}
\sigma=f'_{w}\,D'_{w}\label{A6}
\end{equation}
where the factor $f'_{w}$ is the fraction of nodes that are associated
with the percolating cluster of conductor bonds, and $D'_{w}$ is
obtained from walks over that percolating cluster. Unfortunately,
it is not obvious how to obtain an algebraic expression for $f'_{w}$
in this case of an infinite system having fraction $q<1$ of conductor
bonds. However, an \textit{asymptotic} expression for $f'_{w}$, applicable
to the incipient infinite cluster of conductor bonds, is derived as
follows.

It is reasonable to assume the critical behavior $f'_{w}\sim\left(f_{w}-f_{w}^{(c)}\right)^{\gamma}$
where $f_{w}^{(c)}$ is the value of $f_{w}$ for the system at the
bond percolation threshold $q=q_{c}$. Note that $f_{0}=\left(1-q\right)^{n}$
is the fraction of nodes for which all $n$ attached bonds are insulators
(for example, $n=2d$ for square and cubic networks). Then
\begin{equation}
f_{w}=1-f_{0}=1-\left(1-q\right)^{n}\label{A7}
\end{equation}
so that
\begin{equation}
f_{w}-f_{w}^{(c)}=-\left(1-q\right)^{n}+\left[\left(1-q\right)+\epsilon\right]^{n}\label{A8}
\end{equation}
where $\epsilon=q-q_{c}$. This last expression allows use of the
approximation (Eq. 3.5.8 from Ref. \citep{Handbook})
\begin{equation}
\left(a+\epsilon\right)^{n}\approx a^{n}+na^{n-1}\epsilon\label{A9}
\end{equation}
for $\epsilon\ll a$. In the case that $\epsilon\rightarrow0$,
\begin{align}
f_{w}-f_{w}^{(c)} & \sim-\left(1-q\right)^{n}+\left(1-q\right)^{n}+n\left(1-q\right)^{n-1}\epsilon\nonumber \\
 & =n\left(1-q\right)^{n-1}\epsilon.\label{A10}
\end{align}
Thus $f'_{w}\sim\epsilon^{\gamma}=\left(q-q_{c}\right)^{\gamma}$.
Further, the exponent $\gamma$ must equal $\beta$ in order that
Eq. (\ref{eq18})\textemdash the relation between critical exponents\textemdash is
preserved, in accordance with universality. Thus
\begin{equation}
f'_{w}\sim\left(q-q_{c}\right)^{\beta}\label{A11}
\end{equation}
for systems at the bond percolation threshold $q=q_{c}$.

Note that \textit{numerical} values for $f'_{w}$ in the general case
$q>q_{c}$ can be obtained in the course of calculations involving
walks over the percolating bond cluster. The method is similar to
that used to obtain the fraction $p'$ of sites that comprise the
percolating site cluster, described in Subsec. D of Sec. V.

Consider an infinite, regular network comprised of nodes and ``undefined''
bonds. Then creation of a conductor-bond cluster begins with a ``seed''
node, at which a walker resides. The subsequent behavior of the walker
is dictated by the variable residence time algorithm. When a node
is first visited, any undefined bonds attached to it are converted
to conductor (with probability $q$) or insulator. If conductor, the
newly connected node becomes ``unvisited'' (unless it's already
``visited''), signifying that while it hasn't (yet) been visited,
it is accessible to the walker and so is part of the conductor-bond
cluster. Thus at the end of walk time $t$, the conductor-bond cluster
includes ``visited'' and (possibly) ``unvisited'' nodes. A \textit{finite}
cluster has no ``unvisited'' nodes (which indicates the cluster
is completely confined by insulator bonds). Otherwise the conductor-bond
cluster is considered to be \textit{infinite}, meaning that the cluster
spans the system of size $R(t)$.

Note that $N$ seed nodes will generate $N_{\mathrm{fin}}$ finite
clusters, $N_{\mathrm{inf}}$ infinite (percolating) clusters, and
$N_{0}$ inactive nodes: $N=N_{\mathrm{fin}}+N_{\mathrm{inf}}+N_{0}$.
Thus
\begin{equation}
f'_{w}=f_{w}\left(\frac{N_{\mathrm{inf}}}{N_{\mathrm{inf}}+N_{\mathrm{fin}}}\right)=\frac{N_{\mathrm{inf}}}{N}\label{A12}
\end{equation}
is the fraction of nodes associated with the percolating cluster of
conductor bonds, for a system of size $\left\langle R(t)\right\rangle $.


\begin{thebibliography}{99}
\bibitem{SA}D. Stauffer and A. Aharony, \textit{Introduction to Percolation
Theory}, revised 2nd. ed. (Taylor \& Francis, London, 1994).

\bibitem{Sahimi}M. Sahimi, \textit{Applications of Percolation Theory}
(Taylor \& Francis, London, 1994).

\bibitem{Straley}J. P. Straley, Critical exponents for the conductivity
of random resistor lattices, Phys. Rev. B \textbf{15} (12), 5733-7
(1977).

\bibitem{VS99}C. DeW. Van Siclen, Walker diffusion method for calculation
of transport properties of composite materials, Phys. Rev. E \textbf{59}
(3), 2804-7 (1999).

\bibitem{VS99a}C. DeW. Van Siclen, Anomalous walker diffusion through
composite systems, J. Phys. A: Math. Gen. \textbf{32}, 5763-71 (1999).

\bibitem{VS03}C. DeW. Van Siclen, Effective scalar properties of
the critical region in functionally graded materials, Physica A \textbf{322},
5-12 (2003).

\bibitem{Wang}J. Wang, Z. Zhou, W. Zhang, T. M. Garoni, and Y. Deng,
Bond and site percolation in three dimensions, Phys. Rev. E \textbf{87},
052107 (2013). 

\bibitem{VS02}C. DeW. Van Siclen, Walker diffusion method for calculation
of transport properties of finite composite systems, Phys. Rev. E
\textbf{65}, 026144 (2002).

\bibitem{SierTri}C. DeW. Van Siclen, Conductivity properties of the
Sierpinski triangle, e-print arXiv:1710.06346 (2020). {[}Available
at https://arxiv.org/abs/1710.06346{]}

\bibitem{Handbook}M. Abramowitz and I. A. Stegun, eds., \textit{Handbook
of Mathematical Functions} (Dover, New York, 1965).
\end{thebibliography}
\end{document}